# Temperature Mapping in Urban Biomes Using an Infrared Thermometer: An Investigative Approach to Physics Education

Welington Fabrício dos Santos Costa[1], Jesuíno Alves Martins Júnior[1], André Flávio Gonçalves Silva[2], Rosivaldo Xavier da Silva[3], Ariel Nonato Almeida de Abreu Silva[1,*]

[1] Federal University of Maranhão, Coordination of the Bachelor's Degree Program in Natural Sciences – Physics, Maranhão, Brazil
[2] Federal University of Maranhão, Coordination of the Degree Program in Rural Education, Maranhão, Brazil
[3] Federal University of Recôncavo of Bahia, Center for Science and Technology in Energy and Sustainability, Bahia, Brazil

E-mail: ariel.nonato@ufma.br and rosivaldo.xs@ufrb.edu.br

**Abstract**

In recent years, the frequency of extreme weather events on Earth has significantly increased. This phenomenon is driven by the intensification of the greenhouse effect caused by anthropogenic activities, which leads to temperature variations in urban environments that affect thermal comfort and quality of life. Given this context, the present study investigates temperature mapping in urban biomes using an infrared thermometer, conducted as part of a hands-on workshop offered during the 21st National Science and Technology Week (SNCT). The initiative involved students from the public school system, and was grounded in physics education, aiming to foster scientific enculturation. Participants engaged in a problem-based learning experience, actively contributing to all stages of the knowledge-construction process. The objective of this study was to examine the relationship between the presence of vegetation and its impact on temperature in urban environments. A qualitative and quantitative methodological approach was adopted, enabling the identification of scientific literacy indicators, such as information sequencing, data organization, logical reasoning, hypothesis formulation, justification, and explanation of observed phenomena. The analysis of the students' statements and activity guidelines provided insights into their critical thinking development. The findings indicated that students developed essential skills for understanding physical and environmental phenomena, effectively linking collected data to scientific concepts, and proposing well-supported interpretations. Moreover, experience reinforced the perception of science as a dynamic and investigative process, fostering curiosity and enhancing students' argumentative abilities. Thus, the workshop proved to be an effective strategy for promoting scientific literacy and engaging participants in the study of environmental impact in urban contexts.

Keywords: urban biomes, scientific literacy, temperature mapping, physics education, thermology, climate change.

# 1. Introduction

Environmental problems have increasingly affected people's lifestyles. The effects of anthropogenic changes on natural ecosystems have fostered the formation of urban climates that are progressively more hostile to human quality of life. Silva et al. (2015) pointed out the following points:

> The urban environment is considered to have the highest expression of the relationship between humans and nature, and the interaction between the two is more harmful to the physical environment. Consequently, certain climatic factors change in urban areas, causing problems related to thermal comfort.

Throughout the 20th century, the world experienced exponential population growth, accompanied by an increase in the number of people living in cities, a consequence of the industrialization of urban centers, which attracted rural populations to these spaces (Júnior & Santos, 2014 and Jatobá et al. 2011). These factors have intensified the impact of global warming, making it increasingly necessary to seek alternatives to mitigate the effects of environmental changes in urban areas.

One of the main concerns addressed in this article is the rise in temperature and the consequent formation of urban heat islands, which leads to increased thermal discomfort and various health problems among the population. The Urban Heat Island (UHI) is a climatic phenomenon in which an urban area exhibits different microclimatic conditions compared to rural or natural areas, resulting in higher air temperatures, lower relative humidity, and alterations in wind speed and rainfall patterns, among other factors, thereby favoring an increase in local temperatures (EPA, 2008).

In this context, this work seeks to relate surface temperature to the presence of vegetation and natural elements, approaching, in a more interactive and dynamic way, concepts covered in the disciplines of Thermodynamics and Physics and Environment within the Natural Sciences Physics Teacher Training Program at the Science Center of Bacabal (CCBa-UFMA). To achieve this, a field-based investigative activity was carried out to understand the issue of rising temperatures in cities and their relationship with the materials constituting urban spaces. This activity involved students from public schools in the Médio Mearim microregion who participated in the 21st National Science and Technology Week (SNCT)[1], held from November 19 to 22, 2024, at the UFMA Science Center in Bacabal.Sample text inserted for illustration.

The development of investigative activities is of paramount importance to the teaching-learning process, providing opportunities for scientific literacy among students. As:

---

[1] 21st National Science and Technology Week – Bacabal Campus

> It seems logical that promoting scientific literacy involves providing space, opportunities, and possibilities.
>
> for students to be introduced to scientific concepts and work with them by investigating problems and building connections between their existing everyday knowledge and the new information provided by schoolwork. Therefore, it is characterized by an intense blending of the school world and the outside world, allowing knowledge from both to be applied across these environments (Sasseron, 2018, p. 22).

Thus, scientific literacy can be conceived of as a teaching process that integrates scientific knowledge with students' everyday experiences. This dynamic and interactive approach enables students to become protagonists in the construction of their own knowledge through practices and experiments based on school content, thus allowing them to become active citizens in society. According to Sasseron (2018, p. 14):

> In Brazil, we find authors who use the expressions 'scientific literacy," scientific enculturation,' and 'scientific literacy' to designate the objective of Science Education aimed at fostering students' civic formation through the mastery and use of scientific knowledge and its implications in various spheres of life.

In this way, investigative approaches prove to be more effective in constructing knowledge because, as highlighted by Araújo and Abib (2003) and Reis and Martins (2015), they allow students to formulate hypotheses, test them, develop critical observations, and explain the characteristics of the issue under investigation, thus becoming protagonists of their own learning. Additionally, educators should encourage students to find satisfaction in learning by connecting physics concepts to natural phenomena and everyday applications (Ndihokubwayo et al., 2022).

Based on this perspective, the aim of this study was to propose an investigative activity in which students would analyze real-world issues by collecting temperature data and producing a temperature map, integrating scientific knowledge with their everyday experiences. Additionally, we sought to understand the impact of vegetation and natural elements on thermal regulation, as well as the effects of the absence of these components and the use of paved materials in urban environments. To this end, field activity was conducted based on the concepts of heat and temperature, enabling the collection of results aligned with reality and fostering reflections on the topic addressed.

Therefore, this study offers a significant contribution to the research on the implementation of green areas and the development of materials with higher heat emissivity for urban paving, aiming at social well-being. By integrating sustainable practices and innovative solutions, it seeks not only to improve the quality of the urban environment but also to promote public health and quality of life. The creation of green spaces and the use of materials that reflect heat can help mitigate the effects of urban heat islands, reduce temperatures, and create a more pleasant and healthier environment for all.

Furthermore, this work also contributes to the promotion of pedagogical practices that make science education more dynamic and closer to student realities. Thus, it demystifies the notion that scientific knowledge is distant from everyday life, highlighting its presence in daily life. This is evidenced by the students' accounts, in which indicators of scientific literacy, such as hypothesis formulation, argumentation, explanation, organization, and information sequencing, clearly emerge. It is evident that even without explicit instruction, students already displayed these indicators, demonstrating that scientific knowledge can become more accessible when approached dynamically and within a contextual framework.

Ideas for science education aimed at promoting scientific literacy are supported by several authors, such as Hurd (1998) and Yore, Bisanz, and Hand *et al*. (2003), who emphasize the need for schools to enable students to understand and acquire knowledge about science and its technologies as essential conditions for becoming citizens with a critical perspective of the contemporary world.

## 2. Methodology

### 1.1 Study area

The study was conducted in the municipality of Bacabal, specifically at the Bacabal Science Center (CCBa) of the Federal University of Maranhão (UFMA), located in Avenida João Alberto, S/N, Bambu neighborhood, in the urban zone of the city. Bacabal is situated in the Central Maranhão mesoregion and the Médio Mearim microregion (Fig. 1), with geographic coordinates of 4°13′30″S and 44°46′48″W (Bacabal, 2025). The municipality has a population of 103,711 inhabitants and a territorial area of 1,656.736 km² bordered by the Mearim River. In socioeconomic terms, Bacabal has a Municipal Human Development Index (MHDI) of 0.651 (IBGE, 2022), reflecting both challenges and potentialities within the regional context.

The region's climate is hot and humid tropical, with an average annual temperature of 27.4°C and an average annual rainfall of 1,549 mm. The dry season occurs between September and December, when the highest temperatures are recorded, whereas the rainy season extends from January to June (Bacabal, 2025).

Bacabal serves as the hub municipality of the Mearim Planning Region, encompassing nine municipalities. It stands out as the main industrial, financial, educational, and service center of the region, attracting people from neighboring cities in search of opportunities and various services.

The municipality has a comprehensive educational structure comprising 93 elementary schools (SEMED, 2021) and 16 high schools (IBGE 2022). In terms of higher and technical education, notable institutions include the Federal University of Maranhão (UFMA), the State University of Maranhão (UEMA), the Federal Institute of Education, Science, and Technology of Maranhão (IFMA), and several other Higher Education Institutions (HEIs). Regarding the quality of education, the municipality presents a Basic Education Development Index (IDEB) of 5.4 with for the early years of elementary school, 5.0 for the final years, and 4.9 for high school (IDEB, 2023). Furthermore, the school enrollment rate for children aged 6 to 14 years is 97.7%, while the literacy rate for the population aged 15 and older is 83.3% (IBGE, 2022).

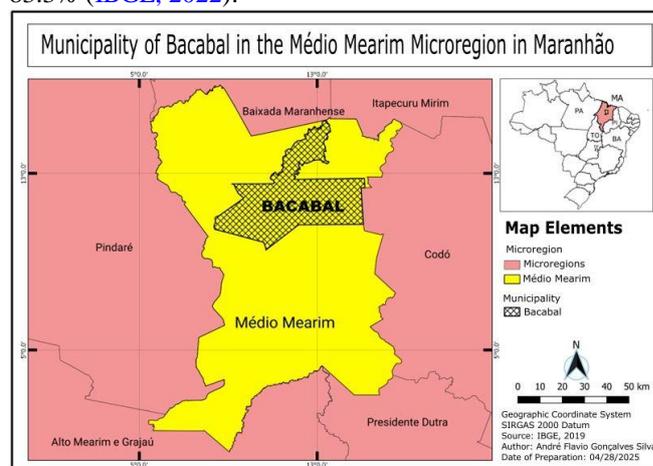

**Figure 1.** Municipality of Bacabal and the Médio Mearim Microregion. Adapted from Silva (2023).

### 1.2 Data collection

The methodology adopted in this study followed a quantitative, qualitative approach. According to Rodrigues, Oliveira, and Santos (2021), the combined use of qualitative and quantitative data in scientific research is essential for a deeper understanding of events, facts, and processes. This approach allows researchers to observe, collect, and compare information and confront their experiences with established scientific knowledge, thereby conferring greater validity to the study. Knechtel (2014, p. 106) complements this view, stating that the qualitative-quantitative methodology *"[...] interprets quantitative information through numerical symbols and qualitative data through observation, participatory interaction, and interpretation of subjects' discourse (semantics)."*

Based on this approach, practical activities with investigative bias were developed, allowing for the integrated collection and analysis of qualitative and quantitative data. These activities were organized within the context of the 21st National Science and Technology Week (SNCT), held at the Bacabal Science Center (CCBa) of the Federal University of Maranhão (UFMA) in Bacabal, from November 19 to 22, 2024, and involved students from the municipal and state public school systems of the Médio Mearim region.

The investigation was structured around a problem-based question: '*How can thermal mapping of urban biomes contribute to identifying heat islands and to sustainable environmental planning aimed at improving thermal comfort in cities?*' To address this question, students were guided to carry out an investigative field practice, recording the temperatures of natural and artificial surfaces in the courtyard of CCBa-UFMA.

Prior to the practical activity, students received instructions regarding the topic and objectives of the workshop. Subsequently, they were organized into five groups of up to three participants to optimize task execution, with each group accompanied by two monitors who provided support throughout the development of the activity.

The monitors responsible for assisting the students underwent prior training conducted by a professor from the Natural Sciences - Physics program, who taught subjects related to the theme of the activity, including Waves and Thermodynamics. This training covered the proper use of measuring equipment, execution of the practical activity, and use of the software employed for data analysis, ensuring qualified guidance and more efficient support for participants.

Afterward, the students were taken to the external courtyard of the campus, where they performed temperature measurements on different types of pavement. During the practice, students were instructed to avoid taking measurements on painted surfaces or those covered with sand, to maintain a standard measurement protocol, and to avoid recording temperatures in the shade cast by their own bodies on the ground, thus ensuring greater accuracy in the results.

Measurements were conducted on three different days, November 19, 21, and 22, 2024, during the morning hours, between 9:00 a.m. and 10:00 a.m. On the first and third days, the sky was predominantly clear and sunny, with little variation in light intensity. However, on the second day, the sky was partially cloudy, reducing the solar incidence for approximately half of the activity duration.

The surface temperature measurements were performed using an infrared thermometer, applying the Celsius scale with an accuracy of ±0.5 ºC. In addition to this device, printed spreadsheets were used for data recording, and a data analysis software was employed to create interpolation matrices for the generation of color maps. The study area, located in the external courtyard in front of the central building, was divided into 5 m² quadrants with measurements taken at the vertices. A total of 12 measurements were made along the east-west axis and six measurements along the north-south axis, forming a rectangle with a total area of 1800 m². To facilitate the practice, it was agreed that students would use their own large steps as the unit of measurement, considering approximately 1 m per step. Thus, measurements were taken every five steps.

The organization of work within the groups followed this structure: one student was responsible for marking the measurement points, another operated the infrared thermometer, and a third recorded the obtained values, with the student responsible for step-counting remaining fixed to maintain measurement standardization.

In the final stage, the collected data were organized and analyzed to create a temperature map. Finally, a session of open-ended questions was conducted, allowing students to express their perceptions about the activity and the knowledge acquired through the practice so that the researchers could analyze the responses and infer indicators of scientific literacy present in the students' answers.

*1.3 Data analysis*

After measuring the temperatures in the courtyard, the students were taken to the Physics Laboratory at CCBa to analyze the collected data. The information obtained during the surface temperature measurements was organized into a matrix using the open-source software SciDAVis (version 2.8), ensuring standardized formatting and logical structuring of the data. This systematic organization allowed for the correct association of each measurement with its corresponding location and environmental conditions, facilitating a comparative analysis among different types of surfaces. Subsequently, the software was used to generate a two-dimensional color-scale temperature map highlighting the spatial distribution of the recorded temperatures across the campus courtyard.

To enhance the analysis and aid students in interpreting the results, images (such as cars, trees, and other elements) previously selected by the monitors were incorporated into the temperature map to represent the physical elements of the investigated environment. This approach enabled a more intuitive visual correlation between different surface types and their respective thermal impacts, contributing to a better understanding of the heat absorption and dissipation processes in urban materials. Furthermore, graphical representation assisted in identifying thermal patterns and fostered discussions on strategies to mitigate the effects of heat islands in built environments.

The findings from this stage provide a basis for reflections on sustainable solutions in urban planning, highlighting the importance of vegetated surfaces and the appropriate use of artificial materials to ensure thermal comfort in cities. In addition, by engaging students in real-world data collection and analysis, the study also reinforces the value of interdisciplinary approaches to address complex socio-environmental challenges.

**Table 1.** Scientific literacy indicators used to analyze students' statements.

| Scientific Literacy Indicator | Description |
|---|---|
| Sequencing of information | *Listing data without a fixed order, used to organize information and facilitate investigations.* |
| Organization of information | *The way data are arranged for better understanding, which can occur at the beginning or during the revisiting of a topic.* |
| Classification of information | *Grouping data based on common characteristics, which may or may not follow a hierarchical structure.* |
| Hypothesis formulation | *Creation of assumptions about a topic, which can take the form of questions or statements.* |
| Hypothesis testing | *Verification of assumptions through experiments or reasoning based on prior knowledge.* |
| Justification | *Presentation of arguments that support a statement, making it more reliable.* |
| Prediction | *Anticipation of a phenomenon or action based on observed facts.* |
| Explanation | *Establishing relationships between information and hypotheses to understand a phenomenon, which can be refined throughout the study.* |

*1.4 Parameters for analysing students' statements*

This study adopted a qualitative approach to assess the levels of scientific literacy demonstrated by students during investigative activities related to urban thermal mapping. Data were collected through open and strategic questioning designed to stimulate reflections on the observed phenomena and capture evidence of the development of scientific thinking.

For the analysis of students' statements, categorization was employed based on the scientific literacy indicators proposed by Sasseron (2018), which allowed for the identification of different levels of students' appropriation of Scientific concepts and investigative practice. These indicators, presented in Table 1, range from the identification of scientific concepts and phenomena to the ability to argue based on evidence and apply knowledge to new contexts.

**3. Results and discussion**

Initially, students participated in a workshop entitled "*Temperature Mapping in Urban Biomes*." During the presentation, fundamental topics for practical activity were discussed, such as the difference between heat and temperature, the formation of heat islands in urban centers, global warming, and its impact on human health and quality. of life. Following this, the materials used throughout the workshop were introduced, including an infrared thermometer, clipboard, pencils, and a data table for recording the collected temperatures. In addition, the use and operation of the infrared thermometer were demonstrated.

Infrared thermometers are devices capable of measuring the heat emitted by an object or the environment without physical contact. They operate on the basis of the principle of thermal radiation, which involves the emission of energy in the form of electromagnetic waves (Grieve *et al*., 2014). Specifically, these thermometers detect infrared radiation emitted by bodies as they heat up. The sensor captures this radiation and converts it into an electrical signal, which is then processed to provide corresponding temperature readings. The amount of energy emitted by the body is directly related to Its thermal state, as described by the Stefan-Boltzmann law, establishes that the radiated power of a black body is proportional to the fourth power of its absolute temperature (Stefan & Boltzmann, 1879).

The accuracy of an infrared thermometer can be influenced by several factors, such as the heat-emission capability of the object, the distance between the device and the target, and the presence of obstructions along the radiation path. Emissivity is a fundamental characteristic of materials that directly influences temperature measurements through thermal radiation (Cosoli *et al*., 2025).

As explained by Incropera and DeWitt (2013), the emissivity of a surface determines the proportion of the thermal energy emitted relative to an ideal black body. This characteristic ranges from 0 to 1 and is affected by various factors, such as the material, surface finish, temperature, observation angle, and wavelength of the emitted radiation. This type of thermometer is a valuable tool because it allows for quick and efficient temperature measurements without the need for direct contact.

Thermal maps were constructed and analyzed based on an area previously determined by workshop monitors, as shown in Fig. 2, which illustrates the arrangement of different elements within the study area. This region features mixed paving, including sections with concrete pavers, smooth concrete flooring, and grass-covered surfaces. Additionally, it contains medium- and large-sized trees, such as the Indian Lilac (*Azadirachta indica*), Java Plum (*Syzygium cumini*), and Yellow Ipe (*Handroanthus albus*), which influence local thermal variation (Singhal *et al*., 2019; Pizzardo *et al*., 2020; Prochnow *et al*., 2010; Qamar *et al*., 2022). Moreover, there is a small canopy at the entrance of the main building, and the Maria Laura Lopes Science Space (ECMLL) is located on the opposite side, creating different shaded areas.

Data collection was carried out by the students under the supervision of the monitors, as illustrated in Figures 3(a) and 3(b). The measurements covered the entire designated area, following a pattern in which recordings were made at the vertices of squares formed by the student responsible for the marking. To optimize the process, the participants were organized into groups of up to three members, distributing different tasks among themselves.

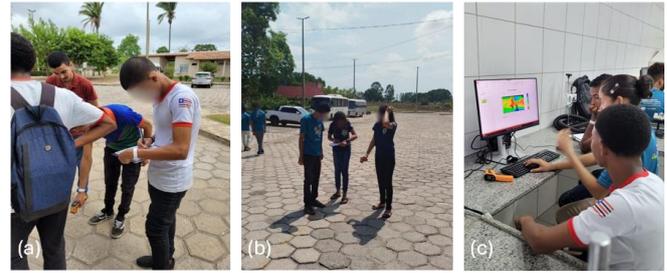

**Figure 3**. Images of students collecting data in the courtyard of the UFMA Bacabal campus: (a) and (b) students recording temperatures in the UFMA courtyard during the investigative activity; (c) students, with the assistance of the monitors, constructing the temperature map based on the collected data.

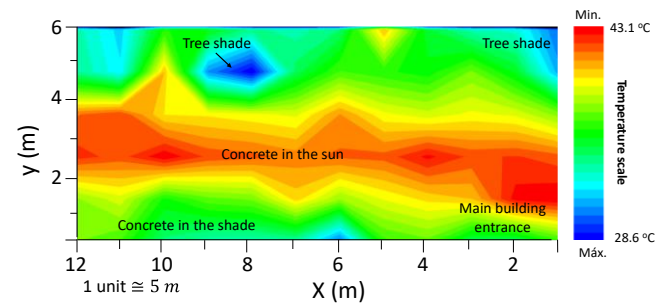

**Figure 4.** Temperature map of the outdoor area in front of the main building of CCBa-UFMA, constructed by the students on 11/19/2024. Adapted from the original.

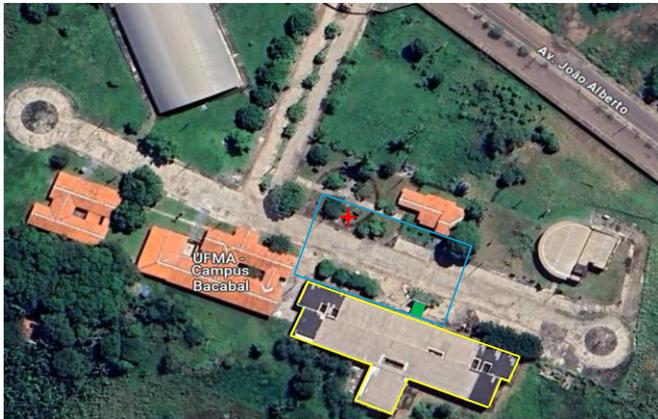

**Figure 2.** Satellite image of UFMA Bacabal Campus. The solid blue line rectangle indicates the designated area for data collection used in the construction of the thermal map. The yellow solid line outlines the main building, and its main entrance is marked by a solid green square. The red symbol highlights a section of the courtyard square that includes vegetation.

Subsequently, the students were encouraged to identify, within the image, the different surface coverings and elements present in the study area. To make the activity more interactive and dynamic, preselected images provided by the monitors were used, and the students were tasked with positioning these images at the corresponding locations where they were found during data collection. This approach, illustrated in Fig. 5, with the thermal map produced by the group monitored on the second day of the practical activity, enabled the association between the thermal map colors and environmental elements, facilitating the understanding of the relationship between surface types and their respective temperatures through a visual analogy.

This map shows that areas under the shade of trees exhibit lower temperatures (bluish and greenish tones), while sun-exposed areas, especially concrete surfaces, record higher temperatures (orange and red tones), reaching a maximum of 43.1°C on this day.

After this stage, the students were taken to the campus physics laboratory to transfer the recorded information into a matrix using the open-source software SciDAVis for the construction of thermal maps (Fig. 3(c)). This process allowed the data to be visualized in a graphical format, facilitating the analysis and interpretation of the thermal patterns in the study area. Fig. 4 presents the color map constructed from the data analysis conducted by one of the groups on November 19, 2024, illustrating the thermal variations recorded in the area on that date.

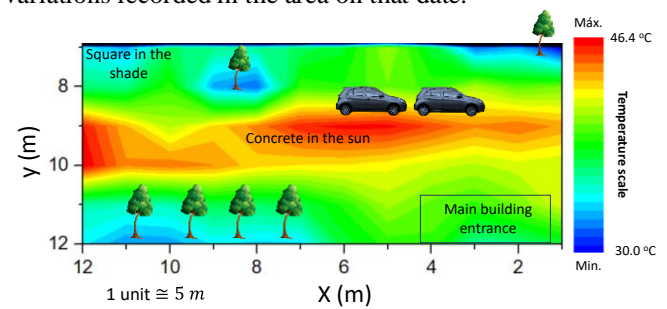

**Figure 5.** A temperature map of the outdoor area in front of the main building of CCBa-UFMA was constructed by the students on 11/21/2024 using an interactive approach. Adapted from the original.

Finally, an open-ended questioning session was conducted on the developed activity and the results obtained, allowing for the identification of the concepts assimilated and knowledge acquired by the students, who were identified as **S1**, **S2**, and **S3**. One of the first questions asked about weather conditions at the time of the activity, and the responses were similar:

**S1:** *The sun was hot, and there was no shade from the clouds, only the shade from the trees.*

**S2:** *When we went outside, it was extremely hot, and the sun was very strong.*

**Teacher:** *Did you enjoy the workshop you participated in?*

**S1:** *Yes. We found it very interesting. Like... because it's a unique experience that we've never had in our daily lives, and... it's something new for us. In previous years, we never had an experience like this — coming to a university and getting to know, like... so deeply, let's say, a subject that is so interesting and important for society.*

**Teacher:** *Do you believe this image represents the reality experienced by our society?*

**S3:** *Yes, very much.*

Scientific literacy is still not fully developed in some schools, as the teaching of Physics and Science often follows a traditional model with little emphasis on investigative and experimental practices. This limitation becomes evident in the statement by student **S1**, who describes the activity as "*a unique experience that we have never practiced in our daily lives,*" highlighting the lack of previous opportunities to experience learning in a more dynamic way. This account demonstrates the need for more active and contextualized methodologies that bring students closer to scientific knowledge in a meaningful way.

Thus, it becomes clear that breaking the paradigm that knowledge can only be acquired in the classroom and transmitted solely by the teacher is an innovative approach, especially since many students still lack access to such experiences. Subsequently, when asked whether there was a difference in the temperature values recorded between shaded and sun-exposed areas, the students presented the following propositions:

**S3**: *Yes. I saw that the temperature changed a lot between the shade and the sun.*

**S2**: *I noticed that too. I already knew it would be different because we can feel it when we walk down the street — when it's really sunny and then we pass under a tree.*

**Teacher**: *So, you're saying that the trees impacted the temperature recorded, is that right?*

**S3**: *Yes. There was even a spot we measured where it was 22°C, then we measured again a bit further ahead and it was 30°C, and later it jumped to 40°C. This change was because of the trees and the heat — where there were trees, the temperature was lower, and where there were none, it was higher.*

**S2**: *Although in some places the grass was low, and in others there were those concrete structures... but still, where there was a tree, the temperature dropped a lot, especially under that big tree in the corner where we finished measuring — that was where it was the coolest.*

**Teacher**: *Great. Regarding the difference found between the temperatures measured under the trees and those found on the concrete, would you be able to explain the relationship between temperature and the material that makes up the elements of the courtyard?*

**S2**: *Well... I think concrete must have something in it that makes it heat up more, like some material they add when making it.*

**S1**: *I think the same. Mainly because concrete covers up everything that's natural — for example, if there's grass somewhere, people go there and cover it with concrete. Then when it rains, it takes longer to cool down.*

**Teacher**: *In the thermal map you built, were you able to easily identify the areas where the trees, concrete pavement, and car parking areas were located?*

**S3**: *Yes, we were.*

**S2**: *I was able to identify them, but not the others... But I remember that the lowest temperature was recorded right there under the trees — it was one of the coolest spots. Where the sun hit directly, like in the middle of that pathway, the temperature was much higher.*

**Teacher:** *Okay. Based on everything you've said, do you believe that urban tree planting is really important for our cities?*

**S1**: *Yes, tree planting is very important both for our air quality and for reducing environmental heat and temperature.*

**S3**: *Yes, it is. But people are cutting everything down — all we see now are houses and asphalt, and nature is being destroyed.*

At this point, we can observe indicators of scientific literacy by analyzing students' statements. In **S2**'s response, the indicator organization of information can be identified, as the student connects his everyday experience to the collected data by stating: "*I noticed that too. I already knew it would be different because we can feel it when we walk down the street — when it's really sunny and then we pass under a tree.*" In this excerpt, he structured his prior knowledge to understand the observed phenomenon, organizing the information in a way that consolidated his interpretation. Additionally, he demonstrated an explanation indicator by attempting to relate the composition of concrete to its ability to absorb heat, seeking to understand the observed thermal phenomenon. This idea is complemented by **S1**, who attempts to justify heat retention in concrete by associating it with the removal of vegetation and its capacity to retain heat for longer periods after rainfall. This demonstrates an effort to understand thermal phenomena based on their perception of the environment.

**S3**, in turn, demonstrates the Classification of information indicator by stating: "*There was an area where we measured 22°C, then a bit further ahead it was 30°C, and later it jumped to 40°C.*" In this excerpt, the student groups the collected data based on their characteristics, comparing temperatures in different areas and the features observed at each location. Additionally, the student also exhibits the Explanation indicator when asserting: "*This change was due to the trees and the heat — where there were trees, the temperature was lower, and where there weren't, it was higher.*" Here, he established a relationship between temperature variation and the presence of trees, seeking to justify the observed difference.

Next, **S1** demonstrates the Justification indicator by arguing: "*Yes, tree planting is very important both for our air quality and for environmental heat.*" The student supported this statement by connecting tree planting with air quality improvement and thermal regulation of the environment.

From this, we can see that even with limited knowledge about the composition of concrete, students **S1** and **S2** agreed with the idea that concrete contains materials that influence its heating and cooling. This is supported by Costa's research (2008, p. 149), which points out that "the specific heat (c) can be considered independent of temperature, with a constant value of c = 1000 J/kg°C for concrete made of siliceous or calcareous aggregates." This means that concrete, in addition to reflecting little solar radiation, has a low specific heat — approximately 0.24 cal/g°C — which causes it to heat up more easily, absorb large amounts of solar radiation energy, and re-emit this energy as heat, thus contributing to the formation of urban heat islands.

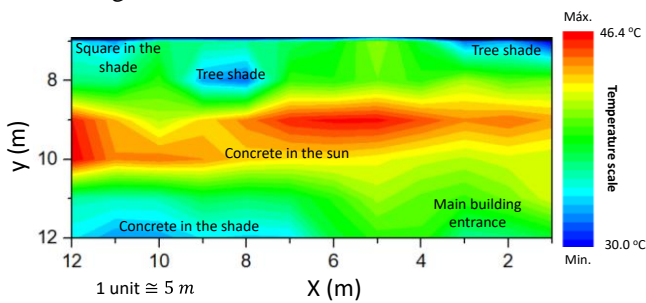

**Figure 6**. Temperature map of the outdoor area in front of the main building of CCBa-UFMA, constructed by the students on 11/21/2024. Adapted from the original.

Based on these accounts, it is evident that, at this initial stage, the students were able to understand the practical activity and construct their own concepts based on the knowledge they produced, aligning with the research conducted by Barbosa (2005), Frota, Jesuíno, and Martins (2018), and Resende (2011), who stated that trees play an important role in regulating environmental temperature and ensuring thermal comfort. Trees not only influence temperature but also affect air humidity, wind speed, and radiation because their canopies reduce the incidence of solar rays on surfaces.

Moreover, it was clear that the activity contributed to the development of students' thoughts, ideas, and concepts, aligning them with the foundations of scientific literacy. By understanding everyday phenomena, such as the relationship between heat and different types of urban paving, and their impact on quality of life, these students moved from a passive attitude toward becoming active agents in scientific knowledge construction.

On the second day of the workshop, another group of students constructed the temperature map shown in Fig. 6 based on the collected data. The study area was the same as that in the previous session, with the only difference being the participation of a new group of students. At this stage, each group consisted of three participants, identified as **S4**, **S5**, and **S6**, to analyze their statements. During this activity, it was observed that sunlight intensity was reduced owing to the presence of clouds in the sky. However, as the activity progressed, the clouds dispersed, allowing the sun to reappear and alter environmental conditions.

After collecting the data and constructing the color map, the students received a new explanation of the relationship between the colors on the scale and measured temperatures. Subsequently, they were asked to identify the elements present in the study area. Then, questioning about the activity and its results began.

**Teacher**: *What did you think about the activity you just completed?*

**S4**: *I thought it was really cool, even though I almost roasted under the sun.*

**S5**: *I liked it too.*

**Teacher**: *I'm glad you enjoyed it. Now, looking at your temperature map, what can you tell me about the colors being presented?*

**S4**: *The blue is where the measurements were lower — the lowest was right there, under the trees. The red is where the temperatures were higher.*

**Teacher**: *Great. You noticed that when we started measuring the temperatures, the sun was covered by clouds, and only later did it come out. Did you feel any difference in the values because of that?*

**S4**: *For sure. A big difference. At first, it was fine, the sun wasn't hot, but once it heated up, we started seeing the temperature go up more and more, and the heat getting stronger.*

**Teacher**: *Were the temperatures lower or higher under the trees?*

**S6**: *Lower.*

**Teacher**: *And when did you notice the temperature started to rise?*

**S5**: *It went up when we... wait, let me remember. Oh! When we went and stopped right there in the middle of the*

*courtyard, and the sun was already hot — right near where the cars pass by. There, we recorded 40°C everywhere the thermometer was placed; the sun got really hot when we got there.*

**Teacher**: *So, what can you say about the presence of vegetation in the area where you measured the temperatures?*

**S4**: *That it helps a lot. The coolest area was right where there was shade, right under the tree where it dropped to 32°C.*

**Teacher**: *And regarding the type of paving in the courtyard, like the concrete areas, what do you have to say? S6: Oh, it gets really hot.*

From these statements, we can identify that student S4 developed the Organization of Information indicator when stating: "*The blue is where the measurements were lower — the lowest was right there, under the trees. The red is where the temperatures were higher*." Here, he demonstrated the ability to structure and interpret the information from the thermal map, correctly associating the colors with the measured temperature values. Additionally, he exhibits the explanation indicator when stating: "*They help a lot. The coolest area was right where there was shade,*" referring to the local vegetation. In this context, he sought to justify the influence of trees on thermal regulation, linking the presence of shade to the reduction in temperature.

Student **S5** developed two indicators: organization and classification of information. In the excerpt "*It went up when we... wait, let me remember. Oh! When we went and stopped right there in the middle of the courtyard, and the sun was already hot — right near where the cars pass by. There, we recorded 40°C everywhere the thermometer was placed*," and he organized the collected data and classified them according to the results obtained, highlighting that the temperature was higher in the middle of the courtyard. Moreover, by stating "*the sun got hot when we got there*," he attempts to explain the temperature increase as a consequence of the sun reappearing, thus demonstrating characteristics of the Explanation indicator.

Subsequently, the questions continued with the students regarding the results and proposals for interventions to improve urban thermal comfort, as presented below.

**Teacher**: *In light of all this, what do you think — does vegetation influence the ambient temperature? Is it really important or irrelevant?*

**S5**: *Man, it's definitely important*.

**S4**: *I think it influences because, where it's hot and there's a tree, it doesn't get as hot because the shade makes the place a little cooler*.

**Teacher**: *So, vegetation does influence it, right?!*

**S5**: *It does*.

**Teacher**: *Lately, there's been a lot of talk about heat waves. Why do you think this is happening?*

**S6**: *Because of the fires, right? People are still burning a lot, cutting down forests, and then there are no more trees and no more coolness.*

**Teacher:** *Is the city where you live hot?*

**S4**: *Yeah, way too hot.*

**Teacher**: *You know we can't just install a giant air conditioner to cool down the whole city. But what do you think should be done to reduce the impact of high temperatures in your city?*

**S4**: *Plant a lot of trees, but now they're just cutting them down. If they keep cutting and burning, the heat will only get worse.*

The Explanation indicator is evident in **S4**'s statement, "*Where there's a tree, it doesn't get as hot because the shade makes the place a little cooler*." In this excerpt, the student seeks to explain the relationship between the presence of vegetation and thermal variation, and associating shade with temperature reduction.

Additionally, the student also demonstrates the prediction indicator by stating: "*If they keep cutting and burning, the heat will only get worse*." Here, he anticipates the consequences of continued deforestation, predicting an increase in temperature based on lived experiences and observed results. Combined, he presents a solution to the problem by suggesting: "Plant a lot of trees," demonstrating that he understands the relationship between the problems and their causes.

Student **S6**: "*Because of the fires, right? People are still burning a lot, cutting down forests, and then there are no more trees and no more coolness*." exhibits the Justification indicator, as he provides an argument to support the claim that heat waves are related to deforestation and fires, showing scientific development connected to daily living practices.

This can be explained by Oke (1982), who stated that paved surfaces without vegetation cover absorb more heat during the day and radiate it more intensely, contributing to the formation of the urban heat island phenomenon. More recent studies, such as those by Lima *et al.* (2023), have pointed out that the introduction of vegetation into urban areas can significantly reduce local temperatures, promoting a more comfortable environment for the population.

Furthermore, it is evident that students recognized the impact of vegetation on thermal regulation. This is important because it fulfills one of the research objectives: to make the student the protagonist of their own learning, in line with the concept of the flipped classroom advocated by Bergmann and Sams (2018).

> "The classroom revolves around the students, not the teacher. [...] The teacher is present solely to provide specialized feedback. Students were also responsible for completing and presenting their schoolwork. Because a solution guide is also offered, students are motivated to learn rather than simply complete tasks from memory. Furthermore, students are encouraged to seek help from the teacher whenever they need assistance in understanding concepts. The teacher's role in the classroom is to support the students, not to transmit information."

Thus, the role of the school and teacher is to act as intermediaries in the students' knowledge-building process, providing adequate support while always encouraging students' autonomy in formulating concepts and developing their understanding.

Continuing the SNCT activities on the third day of work, students from other schools were received. The procedures followed the same format as in the previous days, including a presentation of the workshop and its objectives. After this introduction, the students were divided into groups of three and sent to the outdoor courtyard for data collection. Finally, they were brought to the laboratory, where they analyzed the collected values and constructed the thermal map presented in Fig. 7.

Following this, an investigative questioning session began to explore the knowledge developed throughout the activity, as expressed by students **S7**, **S8**, and **S9**. When asked about their experience of the activity, the following statements were obtained:

**S7**: *I liked it, I thought it was really cool.*

**S8**: *Me too.*

**Teacher**: *Looking at the map and comparing it with the place where you took the measurements, where did you find it to be hotter and where was it cooler?*

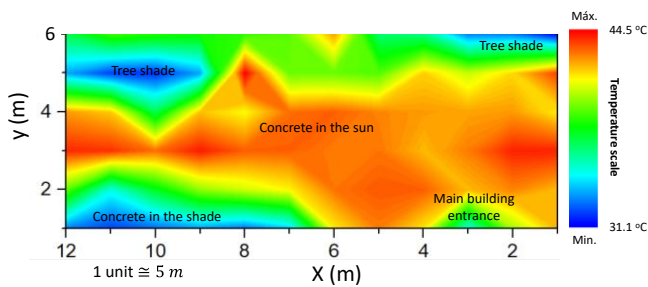

**Figure 7.** Temperature map of the outdoor area in front of the main building of CCBa-UFMA, constructed by the students on 11/22/2024. Adapted from the original.

**S7**: *Where there's vegetation, it's more humid, and it's cooler. Where there isn't, it's hotter.*

**Teacher**: *Right. So, in the hottest areas, there's no vegetation?*

**S7**: *There isn't.*

**Teacher**: *At the school you attend, are there trees?*

**S9**: *There are, but it's in a rural area. But inside the school, it's very hot — we only manage because there's air conditioning and fans. There was a time when we didn't have them, and we were sweating a lot, almost dying from the heat inside.*

**Teacher**: *So, you're saying that a place with trees is better?*

**S9**: *Yes.*

**Teacher**: *On the map, were you able to identify the colors that represent the hottest and coolest areas?*

**S8**: *Red. Red is the hottest, and blue is the coolest.*

**Teacher**: *And outside, where would the red area be?*

**S8**: *On the concrete. Where there are no trees and the cars pass.*

**Teacher**: *Right. And in the grass areas, what happened to the temperature?*

**S7**: *It decreased.*

**Teacher:** *So, do you believe that trees really influence the ambient temperature?*

**S9**: *Yes, I think they do because the shade from the trees makes the place cooler.*

**Teacher:** *And why doesn't the same happen with concrete?*

**S8**: *I don't know how to explain it very well, but I believe it's because concrete is man-made, and things made by humans usually act like that — they're not natural, so they heat up more.*

**S7** demonstrates the Classification of Information indicator by stating: "*Where there's vegetation, it's more humid, and it's cooler. Where there isn't, it's hotter*." In this excerpt, the student grouped locations into two categories based on the relationship between vegetation presence and temperature. Additionally, by saying "*It decreased,*" he verified through observation that temperatures indeed dropped in areas with vegetation, thus demonstrating the Hypothesis Testing indicator.

Student **S8** demonstrated the Sequencing of Information indicator by stating: "*Red. Red is the hottest, and blue is the coolest,*" organizing the data based on the map legend and associating colors with temperature values. He also shows the Organization of Information indicator when saying: "*On the concrete. Where there are no trees and cars pass.*" Here, he spatially structures the information and identifies the hottest locations as paved ones. Finally, he demonstrates the Justification indicator by attempting to explain the thermal difference between natural and man-made areas, saying: "*I don't know how to explain it very well, but I believe it's because concrete is man-made, and things made by humans usually act like that — they're not natural, so they heat up more*." In this passage, he associates artificial material with greater heat absorption, attempting to justify his perception of the phenomenon.

Meanwhile, student **S9** demonstrated characteristics of the explanation indicator by stating: "*Yes, I think they do because the shade from the trees makes the place cooler*," establishing a causal connection between tree shade and temperature reduction. Additionally, by affirming that treed areas are better, he demonstrates the justification indicator, presenting arguments based on his lived experience at school and highlighting the intense heat experienced without cooling equipment. Based on this experience, he reinforced his perception and confirmed the researcher's statement.

Continuing, a brief concept of urban heat islands is presented. In this context, the students were asked if their city was hot.

**S7**: *Yes, very hot. There are places where there's not even shade to go to — in my city, there are hardly any trees, only in front of a school where there's a small square.*

**S8**: *The hottest places are the ones without trees. We feel it a lot when we go downtown.*

**Teacher**: *So, what would you propose to help minimize this problem?*

**S9**: *It would be good to create a project to increase vegetation in the city center. In our city, there are places without any shade. When we go to the street to buy something, we get burned in the sun — if you're not wearing long sleeves, it's hard to bear the heat.*

**S8**: *Exactly, based on what we've seen here, we should do something and plant more trees.*

From these statements, we can identify that in **S7**'s speech, he demonstrates the Justification indicator by stating: "*There are places where there's not even shade to go to — in my city, there are hardly any trees*," providing arguments from daily life experiences to support the previously mentioned information.

Meanwhile, **S8**, by stating: "*The hottest places are the ones without trees*," demonstrates the Classification of Information indicators, establishing a relationship between temperature and vegetation cover, and grouping hotter areas as those lacking trees. Furthermore, along with **S9**, he proposed solutions to mitigate the effects of temperature increase in their cities, demonstrating that they were able to assimilate the research data with the impacts caused by human activities.

The students' propositions are supported by Lima *et al*. (2023), who stated that one of the factors promoting temperature increases in cities is the lack of green spaces and open areas. Moreover, vegetation plays an important role in regulating the temperature in urban centers, and the replacement of these areas with concrete and asphalt decreases the natural cooling capacity of the environment, as unplanned urban infrastructure creates environments that hinder air circulation and heat dissipation.

By constructing a color map, students were able to compare their results and discuss the identified temperature variations, relating them to different types of surfaces and the presence of shaded areas. During this investigation, students engaged in the concepts of heat and temperature, and indirectly established relationships with the heat capacity of the materials. In this context, they understood that certain materials, such as concrete, have lower specific heat, making it easier for them to retain heat. This accumulation occurs throughout the day and is gradually released at night, intensifying the thermal sensation in urban areas both during the day and night.

The results obtained corroborate the idea defended by Lima *et al*. (2023), who emphasized the effectiveness of creating green areas in cities to minimize the impacts of global warming and the formation of urban heat islands, thus promoting better quality of life and providing natural cooling for urban centers.

Based on all the collected information and students' statements, it is clear that changing the traditional classroom context, where the student is merely a passive receiver of knowledge, becomes an innovative approach that contributes to more solid learning. These students confront scientific knowledge with their lived reality, ensuring that knowledge goes hand in hand with critical thinking, thus forming not only individuals who reproduce ready-made information, but also true producers of factual knowledge.

## 4. Conclusions

This investigative study, conducted with students from municipal and state public schools in Maranhão, aimed to establish the relationship between the presence of vegetation and its impact on the temperature in urban environments. To achieve this, students analyzed the variation in surface temperatures on the UFMA campus in the municipality of Bacabal through the construction of thermal maps. The results showed that shaded and vegetated areas presented lower temperatures, while paved surfaces exposed to direct sunlight had higher values, as concrete, owing to its low specific heat, proved to be one of the main contributors to the accumulation and retention of heat throughout the day, thus contributing to the phenomenon of urban heat islands and highlighting the fundamental role of vegetation in urban thermal regulation.

Beyond scientific findings, the project had a significant impact on education, contributing to students' scientific literacy. By investigating everyday physical phenomena, students deepened their understanding of concepts, such as thermal absorption and heat transfer, demonstrating the potential of investigative methodologies to make learning more meaningful. The teaching proposal, based on a problem-posing approach, allowed students to play an active role in several stages of scientific knowledge construction. The teaching-learning relationships were qualitatively assessed based on scientific literacy indicators identified in the students' narratives, such as the classification and organization of information, hypothesis testing, justification, and explanation.

Thus, the investigative approach not only reinforced the importance of sustainable practices but also encouraged students to critically analyze the impact of microclimatic changes and propose interventions to improve the environmental quality of urban spaces. It also fostered a greater environment of engagement and exchange of ideas between students and between students and teachers.

Despite the contributions of this study, several limitations must be considered. This research focused exclusively on surface temperature analysis during a specific period of the year, without accounting for variables such as soil moisture, precipitation, wind speed, and seasonal variations. Future studies could broaden this approach by investigating the impact of different plant species on thermal regulation, the effectiveness of alternative paving materials in mitigating stored heat, and the feasibility of sustainable solutions that reconcile the demands of urban centers with population well-being and environmental preservation.

It is also highly valuable for other groups of teachers and students from different regions to develop similar projects to contribute additional perspectives and to identify different perceptions, challenges, opportunities, and realities. As a challenge, it is proposed that this type of activity be further developed through an interdisciplinary approach that involves teachers from diverse academic backgrounds.

**Data availability statement**

All data supporting the findings of this study are included in the article (and any supplementary files).

**Acknowledgements**

The authors would like to thank the Bacabal Science Center (CCBa) of the Federal University of Maranhão (UFMA), the National Council for Scientific and Technological Development (CNPq) for financial support of the 21st National Week of Science and Technology, and the Coordination for the Improvement of Higher Education Personnel (Capes).